\documentclass{article}
\usepackage[accepted]{icml2006}
\usepackage{graphicx}
\usepackage{amsmath}
\usepackage{amsfonts}
\usepackage{amssymb}

\newcommand{\e}{\mathrm{e}}

\icmltitlerunning{Structural Inference of Hierarchies in Networks}

\begin{document} 

\twocolumn[
\icmltitle{Structural Inference of Hierarchies in Networks}
 
%
%
%

\icmlauthor{Aaron Clauset}{aaron@cs.unm.edu}
\icmladdress{Department of Computer Science,
	University of New Mexico, 
         Albuquerque, NM 87131 USA}
\icmlauthor{Cristopher Moore}{moore@cs.unm.edu}
\icmladdress{Departments of Computer Science and Physics and Astronomy,
	University of New Mexico, 
         Albuquerque, NM 87131 USA}
\icmlauthor{Mark Newman}{mejn@umich.edu}
\icmladdress{Department of Physics and Center for the Study of Complex Systems,
University of Michigan, Ann Arbor, MI 48109 USA}

\vskip 0.3in
]

\begin{abstract} 
One property of networks that has received comparatively little attention is hierarchy, i.e., the property of having vertices that cluster together in groups, which then join to form groups of groups, and so forth, up through all levels of organization in the network.  Here, we give a precise definition of hierarchical structure, give a generic model for generating arbitrary hierarchical structure in a random graph, and describe a statistically principled way to learn the set of hierarchical features that most plausibly explain a particular real-world network.  By applying this approach to two example networks, we demonstrate its advantages for the interpretation of network data, the annotation of graphs with edge, vertex and community properties, and the generation of generic null models for further hypothesis testing.
\end{abstract} 

%
%

\section{Introduction}
Networks or graphs provide a useful mathematical representation of a broad variety of complex systems, from the World Wide Web and the Internet to social, biochemical, and ecological systems.  The last decade has seen a surge of interest across the sciences in the study of networks, including both empirical studies of particular networked systems and the development of new techniques and models for their analysis and interpretation~\cite{albert:barabasi:2002,newman:2003}.

Within the mathematical sciences, researchers have focused on the statistical characterization of network structure, and, at times, on producing descriptive generative mechanisms of simple structures.  This approach, in which scientists have focused on statistical summaries of network structure, such as path lengths~\cite{watts:strogatz:1998,kleinberg:2000}, degree distributions~\cite{barabasi:albert:1999}, and correlation coefficients~\cite{newman:2002}, stands in contrast with, for example, the work on networks in the social and biological sciences, where the focus is instead on the properties of individual vertices or groups.  More recently, researchers in both areas have become more interested in the global organization of networks~\cite{soderberg:2002,wasserman:robins:2005}.

One property of real-world networks that has received comparatively little attention is that of \emph{hierarchy}, i.e., the observation that networks often have a fractal-like structure in which vertices cluster together into groups that then join to form groups of groups, and so forth, from the lowest levels of organization up to the level of the entire network.  In this paper, we offer a precise definition of the notion of hierarchy in networks and give a generic model for generating networks with arbitrary hierarchical structure.  We then describe an approach for learning such models from real network data, based on maximum likelihood methods and Markov chain Monte Carlo sampling.  In addition to inferring global structure from graph data, our method allows the researcher to annotate a graph with community structure, edge strength, and vertex affiliation information.

At its heart, our method works by sampling hierarchical structures with probability proportional to the likelihood with which they produce the input graph.  This allows us to contemplate the ensemble of random graphs that are statistically similar to the original graph, and, through it, to measure various average network properties in manner reminiscent of Bayesian model averaging.  In particular, we can
\begin{enumerate}
\item search for the maximum likelihood hierarchical model of a particular graph, which can then be used as a \emph{null model} for further hypothesis testing,
\item derive a consensus hierarchical structure from the ensemble of sampled models, where hierarchical features are weighted by their likelihood, and
\item annotate an edge, or the absence of an edge, as ``surprising'' to the extent that it occurs with low probability in the ensemble.
\end{enumerate}
To our knowledge, this method is the only one that offers such information about a network.  Moreover, this information can easily be represented in a human-readable format, providing a compact visualization of important organizational features of the network, which will be a useful tool for practitioners in generating new hypotheses about the organization of networks.

\section{Hierarchical Structures}
The idea of hierarchical structure in networks is not new; sociologists, among others, have considered the idea since the 1970s.  For instance, the method known as \emph{hierarchical clustering} groups vertices in networks by aggregating them iteratively in a hierarchical fashion~\cite{wasserman:faust:1994}.  However, it is not clear that the hierarchical structures produced by these and other popular methods are unbiased, as is also the case for the hierarchical clustering algorithms of machine learning~\cite{hastie:tibshirani:friendman:2001}.  That is, it is not clear to what degree these structures reflect the true structure of the network, and to what degree they are artifacts of the algorithm itself. This conflation of intrinsic network properties with features of the algorithms used to infer them is unfortunate, and we specifically seek to address this problem here.

\begin{figure}[t]
\begin{center}
\includegraphics[scale=0.4]{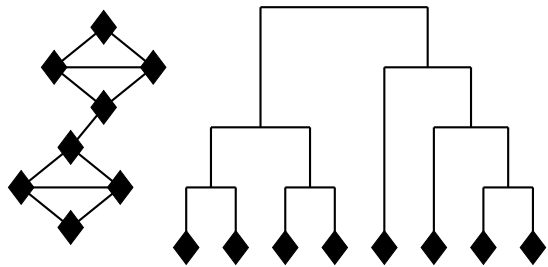} 
\caption{A small network and one possible hierarchical organization of its nodes, drawn as a dendrogram. }
\label{fig:example}
\end{center}
\end{figure}

A hierarchical network, as considered here, is one that divides naturally into groups \emph{and} these groups themselves divide into subgroups, and so on until we reach the level of individual vertices.  Such structure is most often represented as a tree or \emph{dendrogram}, as shown, for example, in Figure~\ref{fig:example}.  We formalize this notion precisely in the following way.  Let $G$ be a graph with $n$ vertices.  A hierarchical organization of $G$ is a rooted binary tree whose leaves are the graph vertices and whose internal (i.e.,~non-leaf) nodes indicate the hierarchical relationships among the leaves.  We denote such an organization by \mbox{$\mathcal{D}=\{D_{1},D_{2},\dots,D_{n-1}\}$}, where each $D_{i}$ is an internal node, and every node-pair $(u,v)$ is associated with a unique~$D_{i}$, their lowest common ancestor in the tree.  In this way, $\mathcal{D}$ partitions the edges of~$G$.

\begin{figure}[t]
\begin{center}
\includegraphics[scale=0.4]{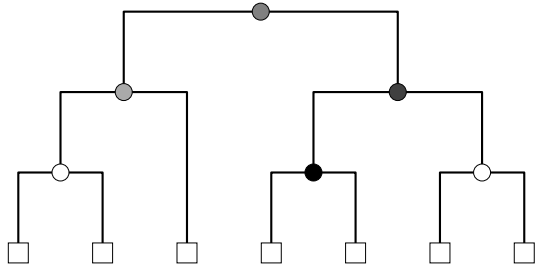}
\caption{An example hierarchical model
$\mathcal{H}(\mathcal{D},\vec{\theta})$, showing a hierarchy among seven graph nodes and the Bernoulli trial parameter $\theta_{i}$ (shown as a gray-scale value) for each group of edges~$D_{i}$.}
\label{fig:dendro}
\end{center}
\end{figure}

\section{A Random Graph Model of Hierarchical Organization}
We now give a simple model $\mathcal{H}(\mathcal{D},\vec{\theta})$ of the hierarchical organization of a network.  Our primary assumption is that the edges of $G$ exist independently but with a probability that is not identically distributed.  One may think of this model as a variation on the classical Erd\H{o}s-R\'enyi random graph, where now the probability that an edge $(u,v)$ exists is given by a parameter $\theta_{i}$ associated with~$D_{i}$, the lowest common ancestor of $u,v$ in~$\mathcal{D}$. Figure~\ref{fig:dendro} shows an example model on seven graph vertices.  In this manner, a particular $\mathcal{H}(\mathcal{D},\vec{\theta})$ represents an ensemble of inhomogeneous random graphs, where the inhomogeneities are exactly specified by the topological structure of the dendrogram $\mathcal{D}$ and the corresponding Bernoulli trial parameters~$\vec{\theta}$.  Certainly, one could write down a more complicated model of graph hierarchy.  The model described here, however, is a relatively generic one that is sufficiently powerful to enrich considerably our ability to learn from graph data.

Now we turn to the question of finding the parametrizations of $\mathcal{H}(\mathcal{D},\vec{\theta})$ that most accurately, or rather \emph{most plausibly}, represent the structure that we observe in our real-world graph~$G$.  That is, we want to choose $\mathcal{D}$ and $\vec{\theta}$ such that a graph instance drawn from the ensemble of random graphs represented by $\mathcal{H}(\mathcal{D},\vec{\theta})$ will be statistically similar to~$G$.  If we already have a dendrogram~$\mathcal{D}$, then we may use the method of maximum likelihood~\cite{casella:berger:1990} to estimate the parameters $\vec{\theta}$ that achieve this goal.  Let $E_{i}$ be the number of edges in $G$ that have lowest common ancestor $i$ in~$\mathcal{D}$, and let $L_{i}$ ($R_{i}$) be the number of leaves in the left- (right-) subtree rooted at~$i$.  Then, the maximum likelihood estimator for the corresponding parameter is $\theta_{i} = E_{i} / L_{i}R_{i}$, the fraction of potential edges between the two subtrees of $i$ that actually appear in our data~$G$.  The posterior probability, or likelihood of the model given the data, is then given by
\begin{align}
\mathcal{L_{\mathcal{H}}}(\mathcal{D},\vec{\theta}) = \prod_{i=1}^{n-1} \left(\theta_{i} \right)^{E_{i}} \left(1 - \theta_{i} \right)^{L_{i}R_{i}-E_{i}} \enspace .
\end{align}
While it is easy to find values of $\theta_{i}$ by maximum likelihood for each dendrogram, it is not easy to maximize the resulting likelihood function analytically over the space of all dendrograms.  Instead, therefore, we employ a Markov chain Monte Carlo (MCMC) method to estimate the posterior distribution by sampling from the set of dendrograms with probability proportional to their likelihood.  We note that the number of possible dendrograms with $n$ leaves is super-exponential, growing like \mbox{$(2n-3)!! \approx \sqrt{2}\,(2n)^{n-1}\e^{-n}$} where $!!$ denotes the double factorial.  We find, however, that in practice our MCMC process mixes relatively quickly for networks of up to a few thousand vertices. Finally, to keep our notation concise, we will use $\mathcal{L}_{\mu}$ to denote the likelihood of a particular dendrogram~$\mu$, when calculated as above.

\section{Markov Chain Monte Carlo sampling}
Our Monte Carlo method uses the standard Metropolis-Hastings~\cite{newman:barkema:1999} sampling scheme; we now briefly discuss the ergodicity and detailed balance issues for our particular application.

\begin{figure}[t]
\begin{center}
\includegraphics[scale=0.4]{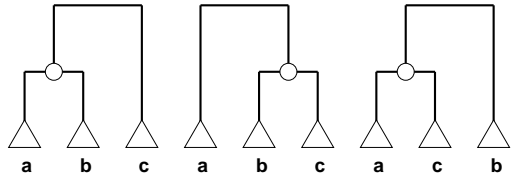}
\caption{Each internal dendrogram node $i$ (circle) has three associated subtrees $a$, $b$, and $c$ (triangles), which together can be in any of three configurations (up to a permutation of the left-right order of subtrees).}
\label{fig:ergo}
\end{center}
\end{figure}

Let $\nu$ denote the current state of the Markov chain, which is a dendrogram~$\mathcal{D}$.  Each internal node $i$ of the dendrogram is associated with three subtrees $a$, $b$, and $c$, where two are its children and one is its sibling---see Figure~\ref{fig:ergo}.  As the figure shows, these subtrees can be in one of the three hierarchical configurations.  To select a candidate state transition $\nu\to\mu$ for our Markov chain, we first choose an internal node uniformly at random and then choose one of its two alternate configurations uniformly at random.  It is then straightforward to show that the ergodicity requirement is satisfied.

Detailed balance is ensured by making the standard Metropolis choice of acceptance probability for our candidate transition: we always accept a transition that yields an increase in likelihood or no change, i.e.,~for which \mbox{$\mathcal{L}_{\mathcal{\mu}} \ge \mathcal{L}_{\mathcal{\nu}}$}; otherwise, we accept a transition that decreases the likelihood with probability equal to the ratio of the respective state likelihoods $\mathcal{L}_{\mu} / \mathcal{L}_{\nu} = \e^{\log\mathcal{L}_{\nu} - \log\mathcal{L}_{\mu}}$.  This Markov chain then generates dendrograms $\mu$ at equilibrium with probabilities proportional to~$\mathcal{L}_{\mu}$.

\section{Mixing Time and Point Estimates}
\label{sec:numeric}
With the formal framework of our method established, we now demonstrate its application to two small, canonical networks: Zachary's karate club~\cite{zachary:1977}, a social network of $n=34$ nodes and $m=78$ edges representing friendship ties among students at a university karate club; and the year 2000 Schedule of NCAA college (American) football games, where nodes represent college football teams and edges connect teams if they played during the 2000 season, where $n=115$ and $m=613$.  Both of these networks have found use as standard tests of clustering algorithms for complex networks~\cite{girvan:newman:2002,radicchi:castellano:cecconi:loreto:parisi:2004,newman:2004} and serve as a useful comparative basis for our methodology.

\begin{figure}[t]
\begin{center}
\includegraphics[scale=0.45]{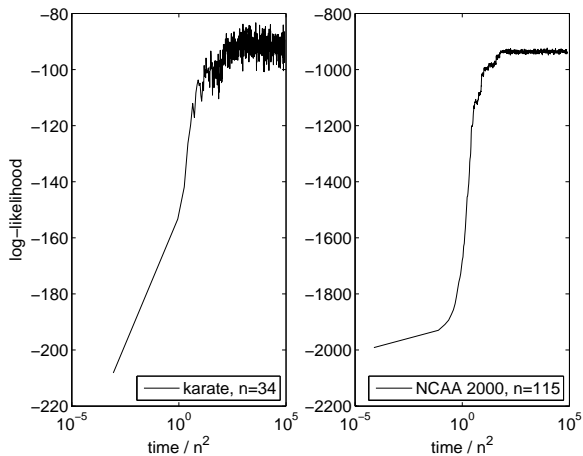} 
\caption{Log-likelihood as a function of the number of MCMC steps, normalized by $n^{2}$, showing rapid convergence to equilibrium.}
\label{fig:converge}
\end{center}
\end{figure}

Figure~\ref{fig:converge} shows the convergence of the MCMC sampling algorithm to the equilibrium region of model space for both networks, where we measure the number of steps normalized by $n^{2}$.  We see that the Markov chain mixes quickly for both networks, and in practice we find that the method works well on networks with up to a few thousands of vertices. Improving the mixing time, so as to apply our method to larger graphs, may be possible by considering state transitions that more dramatically alter the structure of the dendrogram, but we do not consider them here. Additionally, we find that the equilibrium region contains many roughly competitive local maxima, suggesting that any particular maximum likelihood point estimate of the posterior probability is likely to be an overfit of the data.  However, formulating an appropriate penalty function for a more Bayesian approach to the calculation of the posterior probability appears tricky given that it is not clear to how characterize such an overfit. Instead, we here compute average features of the dendrogram over the equilibrium distribution of models to infer the most general hierarchical organization of the network.  This process is described in the following section.

\begin{figure*}[t]
\begin{center}
\begin{tabular}{cc}
\includegraphics[scale=0.45]{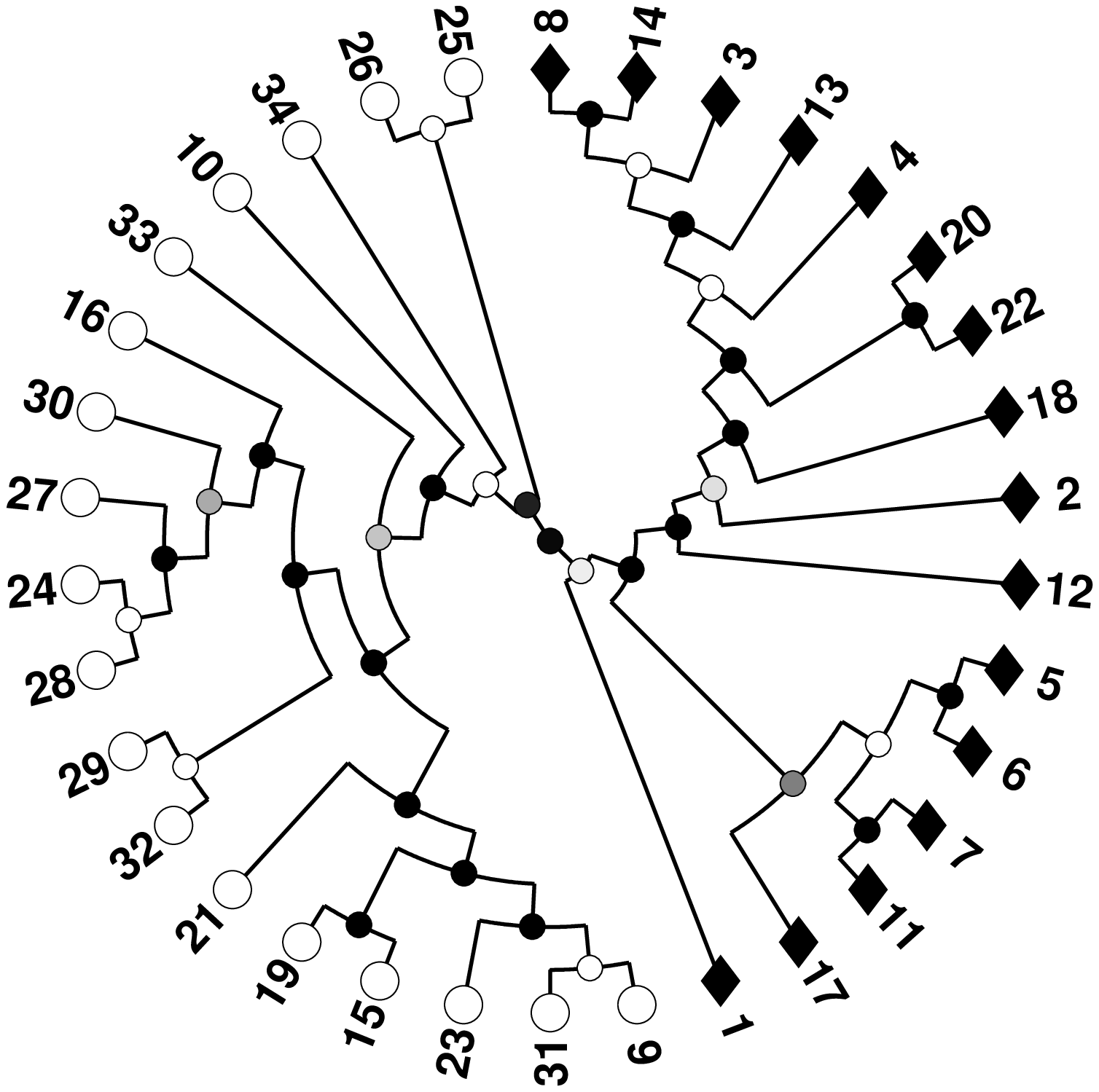} &
\includegraphics[scale=0.45]{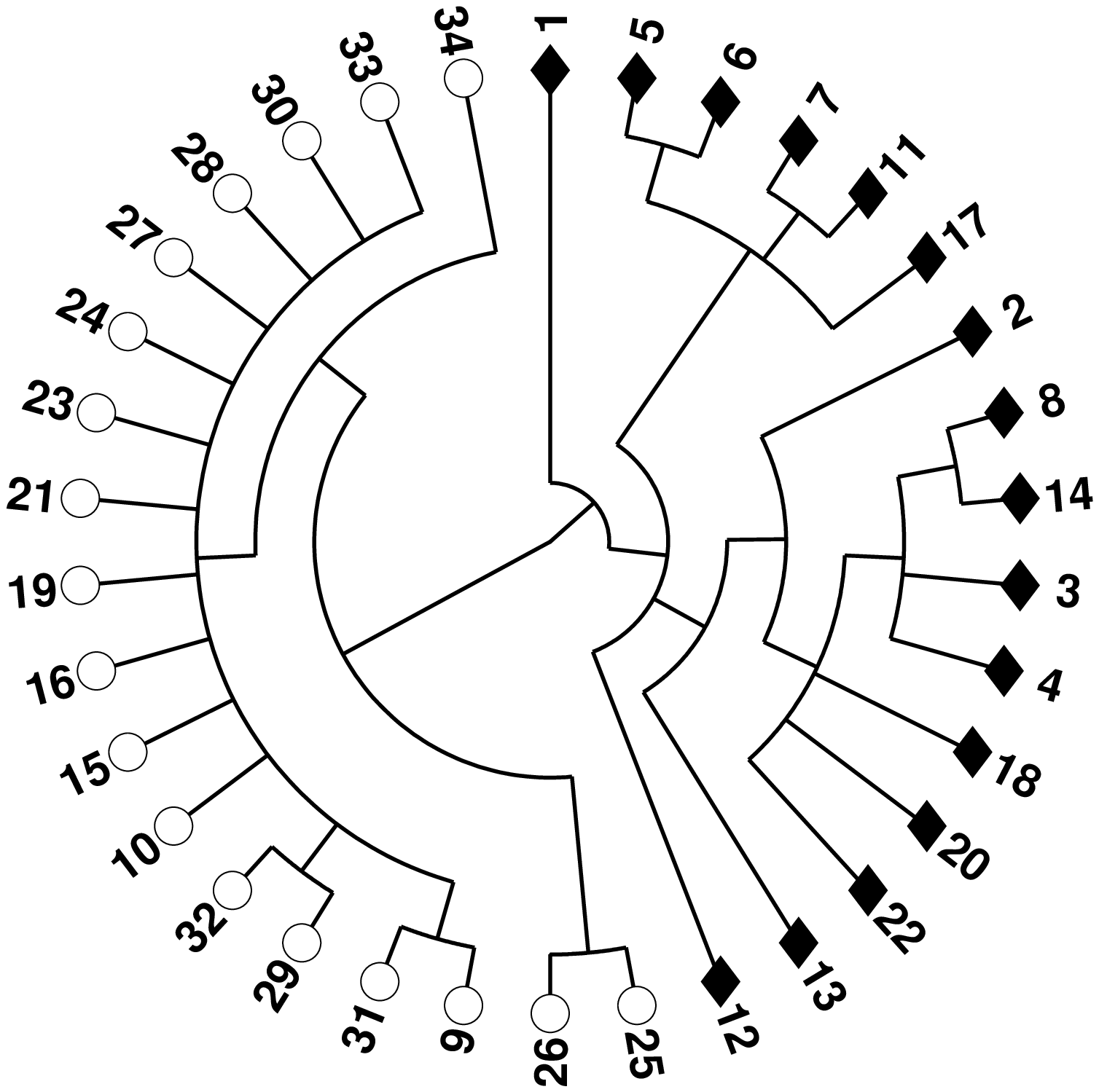} \\
(a) & (b)
\end{tabular}
\caption{Zachary's karate club network: (a) an exemplar maximum likelihood dendrogram with $\log\mathcal{L} = -73.32$, parameters $\theta_{i}$ are shown as gray-scale values, and leaf shapes denote conference affiliation; and (b) the consensus hierarchy sampled at equilibrium.  Leaf shapes are common between (a) and (b), but position varies.}
\label{fig:karate}
\end{center}
\end{figure*}

To give the reader an idea of the kind of dendrograms our method produces, we show instances that correspond to local maxima found during equilibrium sampling for each of our example networks in Figures~\ref{fig:karate} (top) and~\ref{fig:ncaa} (top).  For both networks, we can validate the algorithm's output using known metadata for the nodes.  During Zachary's study of the karate network, for instance, the club split into two groups, centered on the club's instructor and owner (nodes $1$ and $34$ respectively), while in the college football schedule teams are divided into ``conferences'' of 8--12 teams each, with a majority of games being played within conferences. Both networks have previously been shown to exhibit strong community structure~\cite{girvan:newman:2002,radicchi:castellano:cecconi:loreto:parisi:2004}, and our dendrograms reflect this finding, almost always placing leaves with a common label in the same subtree.  In the case of the karate club, in particular, the dendrogram bipartitions the network perfectly according to the known groups.  Many other methods for clustering nodes in graphs have difficulty correctly classifying vertices that lie at the boundary of the clusters; in contrast, our method has no trouble correctly placing these peripheral nodes.

\section{Consensus Hierarchies}
Turning now to the dendrogram sampling itself, we consider three specific structural features, which we average over the set of models explored by the MCMC at equilibrium.  First, we consider the hierarchical relationships themselves, adapting for the purpose the technique of \emph{majority consensus}, which is widely used in the reconstruction of phylogenetic trees~\cite{bryant:2003}.  Briefly, this method takes a collection of trees $\{T_{1},T_{2},\dots,T_{k}\}$ and derives a majority consensus tree~$T_{\rm maj}$ containing only those hierarchical features that have majority weight, where we somehow assign a weight to each tree in the collection. For our purposes, we take the weight of a dendrogram $\mathcal{D}$ simply to be its likelihood~$\mathcal{L}_{D}$, which produces an averaging scheme similar to Bayesian model averaging~\cite{hastie:tibshirani:friendman:2001}.  Once we have tabulated the majority-weight hierarchical features, we use a reconstruction technique to produce the consensus dendrogram.  Note that $T_{\rm maj}$ is always a tree, but is not necessarily strictly binary.

\begin{figure*}[t]
\begin{center}
\begin{tabular}{cc}
\includegraphics[scale=0.5]{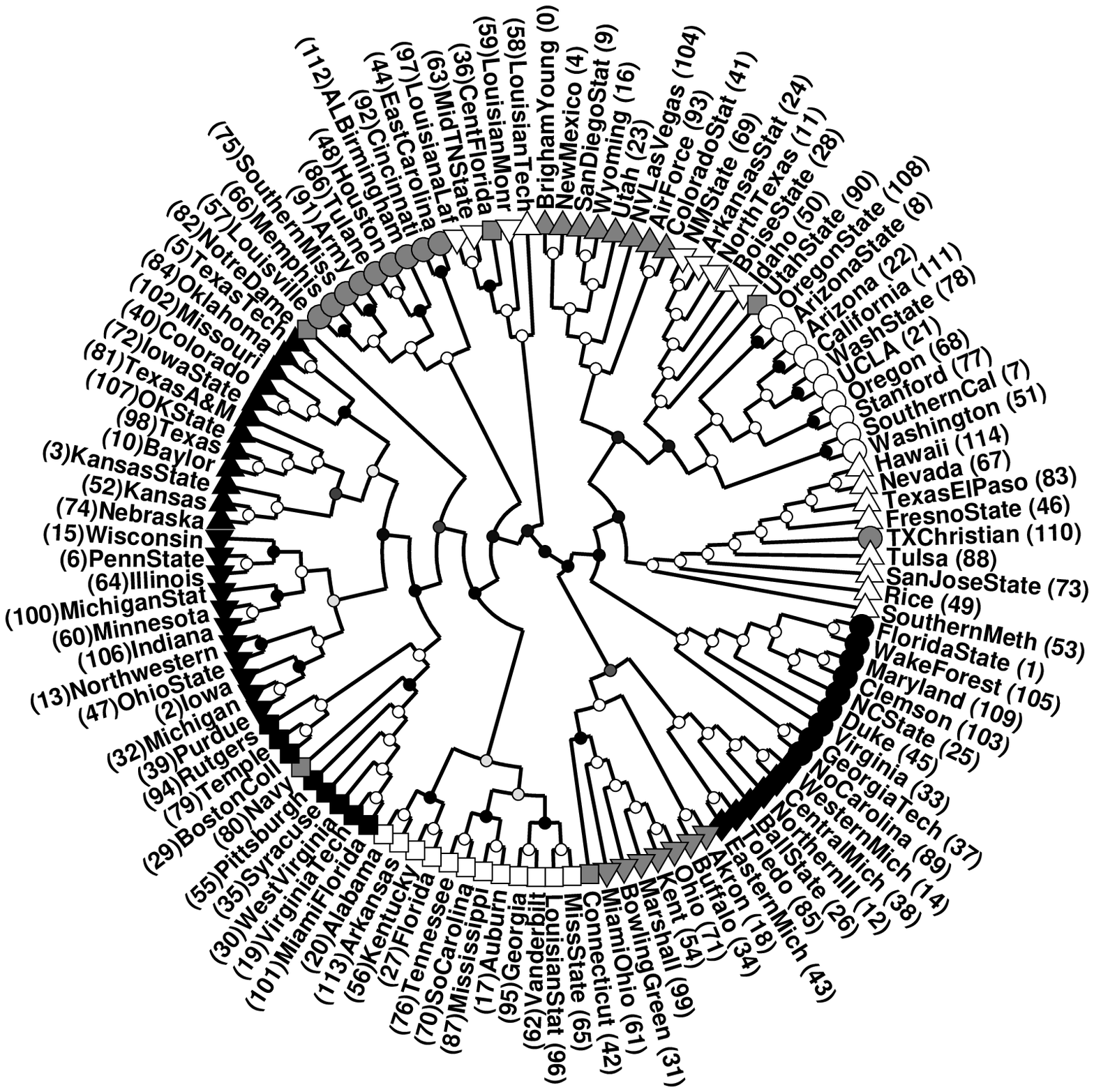} &
\includegraphics[scale=0.5]{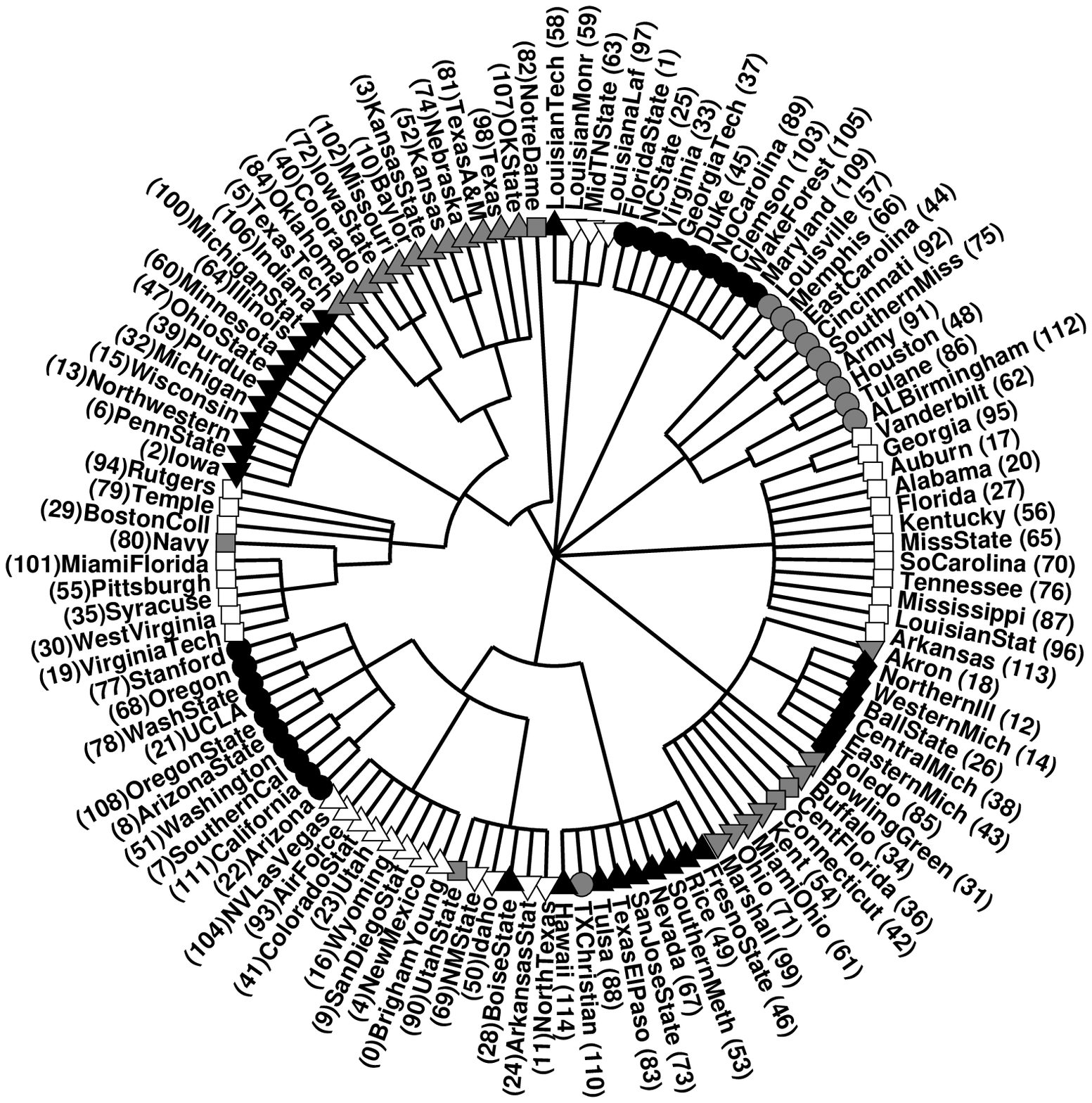} \\
(a) & (b)
\end{tabular}
\caption{The NCAA Schedule 2000 network: (a) an exemplar maximum likelihood dendrogram with $\log\mathcal{L} = -884.2$, parameters $\theta_{i}$ are shown as gray-scale values, and leaf shapes denote conference affiliation; and (b) the consensus hierarchy sampled at equilibrium.  Leaf shapes are common between (a) and (b), but position varies.}
\label{fig:ncaa}
\end{center}
\end{figure*}

The results of applying this process to our example networks are shown in Figures~\ref{fig:karate} (bottom) and~\ref{fig:ncaa} (bottom).  For the karate club network, we observe that the bipartition of the two clusters remains the dominant hierarchical feature after sampling a large number of models at equilibrium, and that much of the particular structure low in the dendrogram shown in Figure~\ref{fig:karate} (top) is eliminated as distracting. Similarly, we observe some coarsening of the hierarchical structure in the NCAA network, as the relationships between individual teams are removed in favor of conference clusterings.

\section{Edge and Node Annotations}
We can also assign majority-weight properties to nodes and edges.  We first describe the former, where we assign a group affiliation to each node.

Given a vertex, we may ask with what likelihood it is placed in a subtree composed primarily of other members of its group (with group membership determined by metadata as in the examples considered here).  In a dendrogram~$\mathcal{D}$, we say that a subtree rooted at some node $i$ encompasses a group~$g$ if both the majority of the descendants of $i$ are members of group~$g$ \emph{and} the majority of members of group $g$ are descendants of~$i$.  We then assign every leaf below $i$ the label of~$g$. We note that there may be some leaves that belong to no group, i.e.,~none of their ancestors simultaneously satisfy both the above requirements, and vertices of this kind get a special no-group label.  Again, by weighting the group-affiliation vote of each dendrogram by its likelihood, we may measure exactly the average probability that a node belongs to its native group's subtree.

Second, we can measure the average probability that an edge exists, by taking the likelihood-weighted average over the sequence of parameters $\theta_{i}$ associated with that edge at equilibrium.

\begin{figure}[t]
\begin{center}
\includegraphics[scale=0.34]{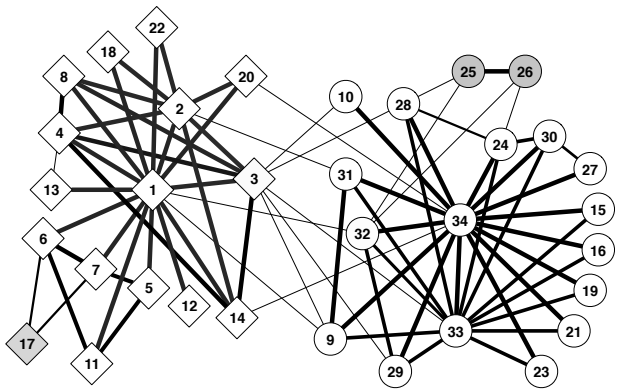}
\caption{An annotated version of the karate club network.  Line thickness for edges is proportional to their average probability of existing, sampled at equilibrium.  Vertices have shapes corresponding to their known group associations, and are shaded according to the sampled weight of their being correctly grouped (see text).}
\label{fig:karate:anno}
\end{center}
\end{figure}

\begin{figure*}[t]
\begin{center}
\includegraphics[scale=0.48]{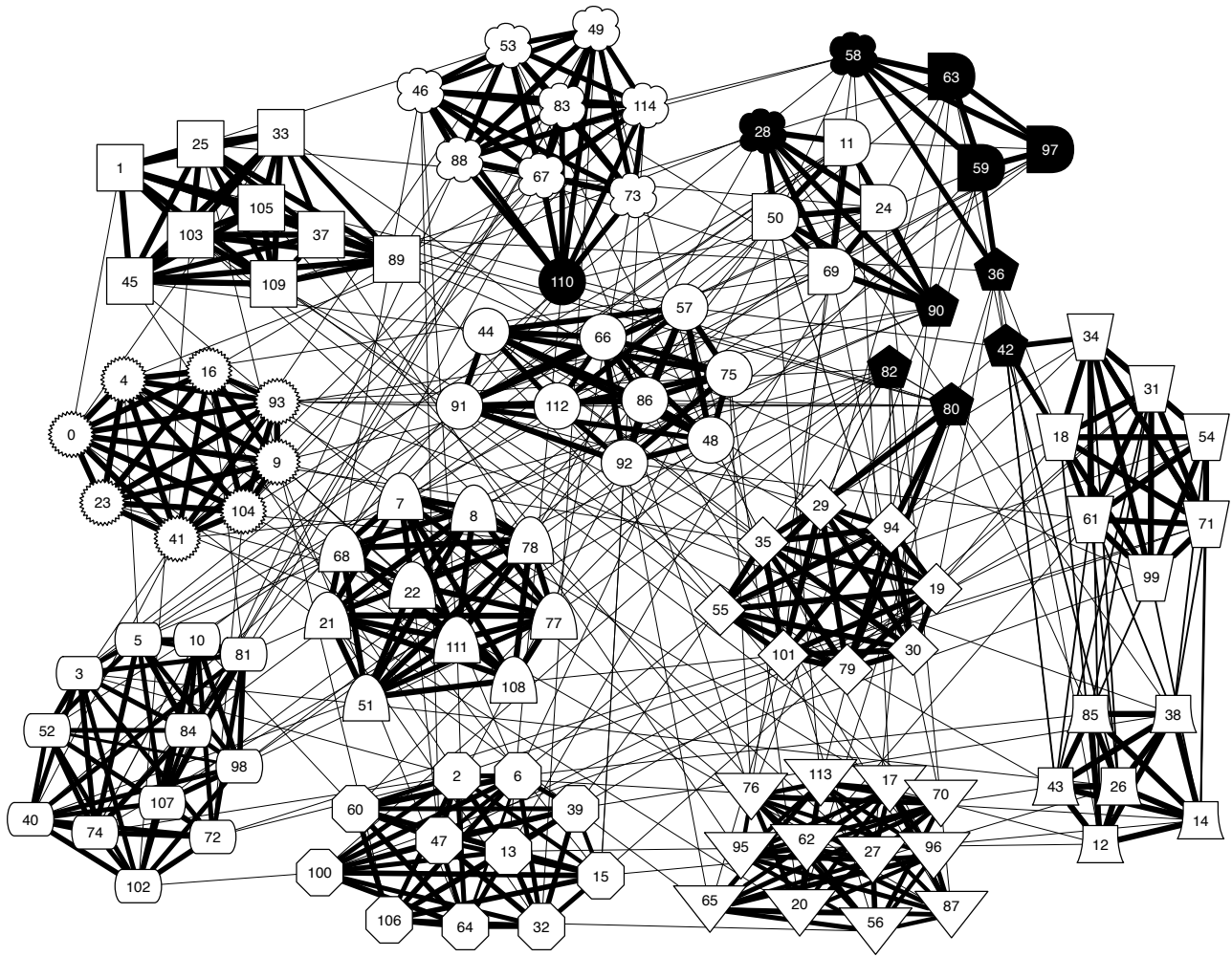} 
\caption{An annotated version of the college football schedule network. Annotations are as in Figure~\ref{fig:karate:anno}.  Note that node shapes here differ from those in Figure~\ref{fig:ncaa}, but numerical indices remain the same.}
\label{fig:ncaa:anno}
\end{center}
\end{figure*}

Estimating these vertex and edge characteristics allows us to annotate the network, highlighting the most plausible features, or the most surprising. Figures~\ref{fig:karate:anno} and~\ref{fig:ncaa:anno} show such annotations for the two example networks, where edge thickness is proportional to average probability, and nodes are shaded proportional to the sampled weight of their native group affiliation (lightest corresponds to highest probability).

For the karate network, the dendrogram sampling both confirms our previous understanding of the network as being composed of two loosely connected groups, and adds additional information.  For instance, node $17$ and the pair $\{25,26\}$ are found to be more loosely bound to their respective groups than other vertices -- a feature that is supported by the average hierarchical structure shown in Figure~\ref{fig:karate} (bottom).  This looseness apparently arises because none of these vertices has a direct connection to the central players $1$ and~$34$, and they are thus connected only secondarily to the cores of their clusters.  Also, our method correctly places vertex $3$ in the cluster surrounding~$1$, a placement with which many other methods have difficulty.

The NCAA network shows similarly suggestive results, with the majority of heavily weighted edges falling within conferences.  Most nodes are strongly placed within their native groups, with a few notable exceptions, such as the independent colleges, vertices~82, 80, 42, 90, and 36, which belong to none of the major conferences.  These teams are typically placed by our method in the conference in which they played the most games.  Although these annotations illustrate interesting aspects of the NCAA network's structure, we leave a thorough analysis of the data for future work.

\section{Discussion and conclusions}
As mentioned in the introduction, we are not the first to study hierarchy in networks.  In addition to persistent interest in the sociology community, a number of authors in physics have recently discussed aspects of hierarchical structure~\cite{girvan:newman:2002,ravasz:somera:mongru:oltavai:barabasi:2002,clauset:newman:moore:2004,salespardo:2006}, although generally via indirect or heuristic means.  A closely related, and much studied, concept is that of community structure in networks~\cite{girvan:newman:2002,radicchi:castellano:cecconi:loreto:parisi:2004,newman:2004,bansal:blum:chawla:2004}\nocite{CSCC04,GA05,DDDA05}. In community structure calculations one attempts to find a natural partition of the network that yields densely connected subgraphs or communities.  Many algorithms for detecting community structure iteratively divide (or agglomerate) groups of vertices to produce a reasonable partition; the sequence of such divisions (or agglomerations) can then be represented as a dendrogram that is often considered to encode some structure of the graph itself.  (Notably, a very recent exception among these community detection heuristics is a method based on maximum likelihood and survey propagation~\cite{hastings:2006}.)

Unfortunately, while these algorithms often produce reasonable looking dendrograms, they have the same fundamental problems as traditional hierarchical clustering algorithms for numeric data~\cite{hastie:tibshirani:friendman:2001}.  That is, it is not clear to what extent the derived hierarchical structures depend on the details of the algorithms used to extract them.  It is also unclear how sensitive they are to small perturbations in the graph, such as the addition or removal of a few edges.  Further, these algorithms typically produce only a single dendrogram and provide no estimate of the form or number of plausible alternative structures.

In contrast to this previous work, our method directly addresses these problems by explicitly fitting a hierarchical structure to the topology of the graph.  We precisely define a general notion of hierarchical structure that is algorithm-independent and we use this definition to develop a random graph model of a hierarchically structured network that we use in a statistical inference context.  By sampling via MCMC the set of dendrogram models that are most likely to generate the observed data, we estimate the posterior distribution over models and, through a scheme akin to Bayesian model averaging, infer a set of features that represent the general organization of the network.  This approach provides a mathematically principled way to learning about hierarchical organization in real-world graphs.  Compared to the previous methods, our approach yields considerable advantages, although at the expense of being more computationally intensive.  For smaller graphs, however, for which the calculations described here are tractable, we believe that the insight provided by our methods makes the extra computational effort very worthwhile.  In future work, we will explore the extension of our methods to larger networks and characterize the errors the technique can produce.

In closing, we note that the method of dendrogram sampling is quite general and could, in principle, be used to annotate any number of other graph features with information gained by model averaging.  We believe that the ability to show which network features are surprising under our model and which are common is genuinely novel and may lead to a better understanding of the inherently stochastic processes that generate much of the network data currently being analyzed by the research community.

%
\section*{Acknowledgments} 
AC thanks Cosma Shalizi and Terran Lane for many stimulating discussions about statistical inference, and Mason Porter for discussions about hierarchy. MEJN thanks Michael Gastner for work on an early version of the model. This work was funded in part by the National Science Foundation under grants PHY--0200909 (AC and CM) and DMS--0405348 (MEJN) and by a grant from the James S. McDonnell Foundation (MEJN).


\end{document}